# Study of space charge force for a laser-accelerated proton beam[*]

Zhu Jungao[1], Zhao Yuan[1], Lai Meifu [1], Gu Yongli [1], Xu Shixiang[2], Zhou Cangtao[1], Lu Haiyang[1]

(1. *Shenzhen Key Laboratory of Ultraintense Laser and Advanced Material Technology*, *Center for Advanced Material Diagnostic Technology*, *College of Engineering Physics*, *Shenzhen Technology University*, *Shenzhen* 518118, *China*;

2. *Shenzhen Key Laboratory of Micro-Nano Photonic Information Technology*, *Key Laboratory of Optoelectronic Devices and Systems of Ministry of Education and Guangdong Province*, *College of Physics and Optoelectronic Engineering*, *Shenzhen University*, *Shenzhen* 518060, *China*)

**Abstract:** Laser accelerators can provide proton beams with unique qualities, such as micron size, picosecond pulse duration and high peak current, and have been demonstrated for various applications and for scientific research purposes. The effect of the space charge force in high peak current beams is strong and raises challenges for application after beam transportation. We performed two-dimensional particle-in-cell simulations and studied the influence of electrons that have velocities close to that of the protons after laser acceleration. We employed ellipsoid models with different charge distributions to estimate the effects of the space charge force. Results demonstrate that space charge will affect beam transmission, and even lead to complete transmission failure if the number of protons per pulse exceeds $10^{10}$. The influence of the space charge force diminishes greatly after 20 ps, which corresponds to approximately 1.2 mm from the target.

Over the past few decades, acceleration using thin solid targets irradiated with high-intensity lasers has developed into a highly promising technology. A reduction in the size and cost of future particle accelerators may be realized by relying on the TV/m electric field gradient that can be established when lasers interact with plasma[1].

Multiple acceleration mechanisms have been explored in laser acceleration, such as Target Normal Sheath Acceleration (TNSA)[2–4], Radiation Pressure Acceleration (RPA)[5–9] and Breakout Afterburner (BOA)[10]. TNSA is the simplest and the most reliable regime so far, driven by the space-charge field generated by hot electrons reaching the rear side of the target. A maximum proton energy of 85 MeV has been achieved using a laser pulse of roughly 100 J per shot in this regime[11]. For laser pulses with high contrast ratios, acceleration efficiencies with nanometer thin foils are higher in the RPA or BOA regimes, and proton beams up to nearly 100 MeV have been demonstrated[12,13].

The unique features of laser-accelerated ion beams, such as small beam size (several microns), short temporal duration (several picoseconds)[14], high brightness, and small emittance, provide important choices for many applications. This technology has great potential for application in isotope production, radiotherapy, fast ignition [15], etc. For fast ignition, proton beams provide a higher energy deposition efficiency than electron beams; the requirements for proton beams are energies ~13 MeV, number of protons ~$10^{16}$, time scale ~100 ps, and spatial scale ~100 μm[16,17]. Ions with energies in the tens of MeV may be utilized for implantation as surface modification of biomaterials or modifying the composition to produce new compounds and nanostructured species[18,19]. Laser-driven proton beams can be used for ultrafast transient process imaging to measure the dynamic evolution of electric fields [20].

The study of warm dense matter (WDM) has important scientific significance in the fields of planetary astrophysics, inertial confinement fusion, high-energy particle beam-matter interactions, among others[21,22]. Particle beams produced by accelerators have many advantages for the production of WDM, such as isochorical heating of solid materials, large volume, no shock wave, and high frequency[23]. Laser-driven proton beams with μm-scale size and ps-scale pulse length have been applied to the study of WDM[24]. The smaller the size of the proton beam and the shorter the pulse length, the more energy is deposited in a certain volume, and the higher the temperature of the WDM. For research, the sample is usually placed within a distance of no more than 1 mm behind the proton beam source to keep the size of the proton beam at tens of μm and the pulse length at tens of ps. As the distance from the laser target is so small, the hot electrons and strong electromagnetic pulses generated by the acceleration process interfere with the analysis of the WDM. In addition, due to the instability of the interaction between the laser and the plasma, the proton beam generated by the acceleration has fluctuations in its energy, spectrum and total charge. Therefore, it is necessary to transmit the particle beam to the application terminal using a beamline, and perform non-destructive diagnosis on the beamline, controlling the energy, energy spread, and charge of the particle beam. With proper beam transport optics, the distribution of the beam at the application terminal will return to the state at the beam source.

The influence of the space charge force in such a system needs to be investigated as an important prerequisite for particle beam transport.

The short beam generated in laser acceleration usually has $10^8 - 10^{10}$ ions, while it is as small as few microns in diameter and as short as tens of picoseconds in pulse length, meaning that the peak current can be of the order of amperes[25]. The repulsive forces due to space charge may cause the beam to expand rapidly and affect the beam quality. References give a space-charge limit of $10^9$ per pulse for laser-accelerated protons[26,27]. For more demanding applications, such as $10^{16}$ protons per pulse for fast ignition, a more detailed study of the effects of space charge is necessary, and steps should be taken to reduce its impact. In addition, the influence of electrons generated during acceleration requires careful study.

A laser plasma accelerator is under construction at Shenzhen Technology University, based on a 5-Hz 200-TW Ti:sapphire laser system using double chirped pulse amplification (CPA) technology. Our aim is to carry out research into the beam transport dynamics in order to further the goal of turning laser accelerators into practical scientific tools.

## 1  PIC simulations of acceleration

The proton beam generated in laser acceleration contains co-moving electrons which have broad energy spectrum in a $4\pi$ solid angle. The electrons that have the same velocities and angles as the protons can counteract the space charge effect before encountering electromagnetic field elements.

In order to study the influence of these co-moving electrons, we performed a series of two-dimensional particle-in-cell (PIC) simulations. In our simulations, the *p*-polarized laser pulses were incident onto different targets—aluminum, plastic (hydrogenated polymer, -($C_5O_2H_8$)-m), diamond-like carbon (DLC)—with various thicknesses, at an incident angle of 30° with respect to the target normal direction, as shown in Fig. 1. The laser has a wavelength of 0.8 μm and peak intensity $I_0 = 2.14 \times 10^{20}$ W/cm², corresponding to a normalized amplitude of $a_0 = 10$. The laser pulse has a Gaussian envelope, i.e., $a = a_0 \exp(-y'^2 / r_0^2)$, where $r_0 = 5$ μm is the laser focal spot radius, with a full-width-half-maximum (FWHM) pulse width of 30 fs. The simulation box size was 40×40 μm and the box had 3200 cells in each direction. There were 64 macroparticles for all species in each cell.

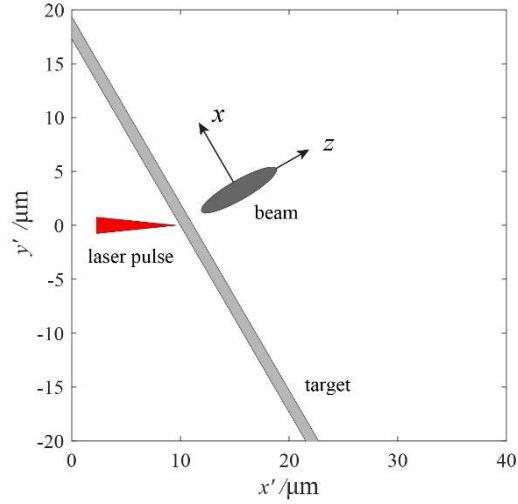

Fig. 1 Schematic of the target and the beam generated in acceleration.

The laser pulse entered from the left boundary along the $z'$ axis, and was situated in the middle of the $y'$ dimension. The targets were placed 10 μm from the left boundary with an angle of 30° with respect to the $y'$ axis. In the simulations with aluminum or plastic targets, a laser pre-pulse is included, which generates a density distribution of $D \times \exp(l / (0.45 \text{ μm}))$ before the target; D is the density of the target; $l$ is the distance from the front surface of the target and within a range of (-4 μm ~ 0). For the ultra-thin foil of DLC, a high contrast ratio of the laser pulse is required and the pre-pulse is removed in these simulations. The thicknesses and densities of the targets are given in Table 1.

Table 1 Proton and electron numbers for simulations with different targets.

|  | Thickness/μm | Density/(g/cm$^3$) | Proton number | Electron number |
| --- | --- | --- | --- | --- |
| Aluminum | 2.5 | 2.7 | $4.22 \times 10^7$ | $7.63 \times 10^{10}$ |
| Polymer | 1.2 | 1.186 | $9.91 \times 10^7$ | $3.87 \times 10^{10}$ |
| DLC | 0.005 | 3 | $2.15 \times 10^7$ | $2.78 \times 10^8$ |

An electron with an energy of 0.011 MeV has the same velocity as a 20-MeV proton. Compared with ion beams with a small energy bandwidth from conventional accelerators, laser-accelerated proton beams are broadband and possess large divergence angles[28,29]. The numbers of these electrons and protons both within an energy spread of ±10% and initial divergence angle of ±50 mrad are shown in Table 1 for simulations with different targets. It can be seen that electron numbers are dominating in all cases.

## 2 Calculations of space charge force

After acceleration, the beam is a mixture of electrons and protons, and transports along the target normal direction. A coordinate system $xyz$ is introduced to describe the expansion of the beam; the $z$ axis is along the target normal direction, as shown in Fig. 1. We set the beam pulse duration to 5 ps. The central energies for protons and electrons were 20 MeV and 0.011 MeV respectively, and both within an energy spread of ±10% and initial divergence angle of ±50 mrad. The initial transverse radius of the beam spot was set to 5 μm. The beam pulse length is about 304 μm, according to the beam pulse duration of 5 ps.

An ellipsoid model is used to compute the space charge force of the beam. The charge distribution is rotationally symmetric about the major axis, which is along the $z$ axis. In our simulations, the semi-minor axis $a$ (along the $x$ axis) and semi-major axis $b$ (along the $z$ axis) are 5 μm and 152 μm respectively at the beginning; $a$ and $b$ keep growing with the expansion of the beam. In the calculation, the time step is 0.01 ps; particles are distributed on grid points with a cell size of $0.01d \times 0.01d \times 0.01d$, ($d = \sqrt[3]{a^2 b}$). Each grid point represents a macroparticle; the charge of macroparticles may be different.

The precise energy spectrum and the initial distribution of the beam may vary under different acceleration conditions. Charge distributions inevitably affect the transverse and longitudinal electric fields, and thereby affect the expansion of the beam. The actual distribution of beam is diverse and difficult to measure directly, and it generally does not exactly fit an ideal model. Moreover, during the expansion process, the distribution of the particle beam may change. Investigating the differences between different models helps to understand the limits of variation in the effect of space charge forces of actual beams. In our simulations, we used three different charge distribution models (linear-descent, waterbag, uniform) for the calculations.

The charge of the beam refers to the total effective charge after neutralizing with electrons. During the drift after acceleration, the total charge does not change, while $a$ and $b$ keep growing and the charge on each grid point changes. The charge distribution density is determined by the position of the grid points. In the uniform distribution, the charge is evenly distributed to each grid point. In the waterbag distribution, the charge of the macroparticle on each grid point is proportional to $(1 - (x^2+y^2)/a^2)(1- z^2/b^2)$. In the linear-descent distribution, the charge on each grid point is proportional to $(1 - \sqrt{x^2 + y^2}/a)(1 - |z|/b)$.

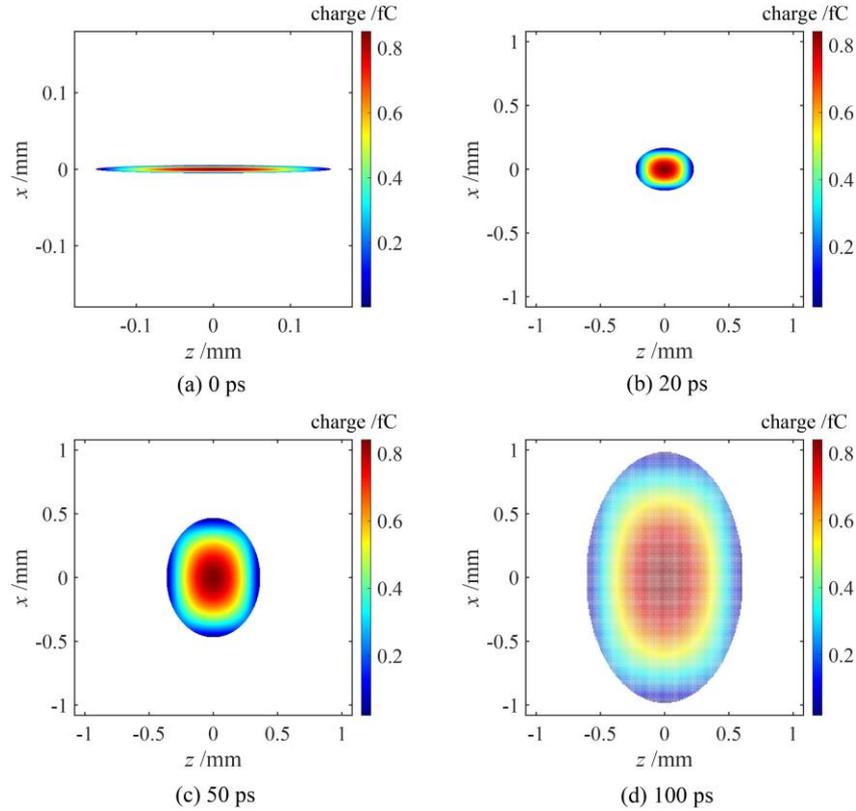

Fig. 2  Beam expansion with a waterbag distribution. Plots show the charge distribution on grid points in the $zx$ plane at (a) 0 ps, (b) 20 ps, (c) 50 ps and (d) 100 ps after the laser interaction with the target.

At each time step of 0.01 ps, the transverse electric field $E_x$ (along the semi-minor axis $a$, as marked in Fig. 2) and longitudinal electric field $E_z$ (along the semi-major axis $b$) are calculated according to the charge on each grid point and considered as fixed for that time step. Then the beam will expand under the influence of the electric fields, energy spread and divergence angle. The proton with the maximum transverse velocity on the outside surface at the semi-minor axis $a$ receives the largest transverse electric force and represents the maximum envelope. The proton at the front of the beam with the maximum longitudinal velocity receives the largest longitudinal electric force and keeps ahead.

As the particle beam expands, the charge density and the effect of the electric field decrease, and the divergence angle and energy spread increase. For example, when the peak current of the beam is 320 A (corresponding to $10^{10}$ protons per pulse), with a waterbag distribution, the charge distribution on grid points in the $zx$ plane at different times after the laser interaction with the target is shown in Fig. 2. The charge is related to the number of macroparticles in the calculation and the cell size; in practice the charge density decreases with the beam expansion.

Figure 3 shows the transverse electric field intensity, longitudinal electric field intensity, divergence angles, beam envelopes (equal to the semi-minor axis $a$), energy spread and longitudinal lengths (twice the semi-major axis $b$) during the expansions within 100 ps, for the three models (linear-descent, waterbag, uniform), with $10^{10}$ protons per pulse. The divergence angle and envelope have the largest increments for the linear-descent distribution; the increases of energy spread and length are largest for the uniform distribution. As the total charge is the same, these results indicate that in the linear-descent distribution, the influence of the transverse electric field is greater, for the increments of divergence angle and envelope are greater; whereas the longitudinal electric field has a greater effect in the uniform distribution, for the increments of energy spread and longitudinal length are greater. The actual situation should be between the results of these two models, and may be close to the results of the waterbag distribution.

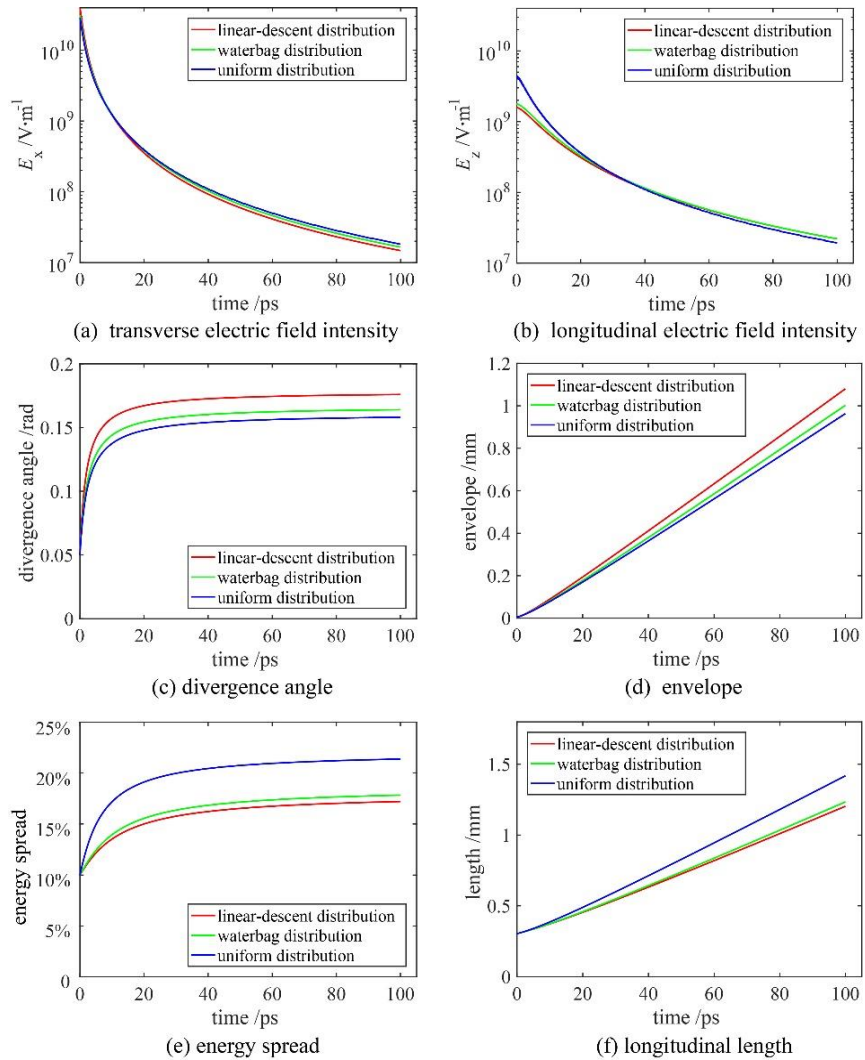

Fig. 3  Influence of space charge in different charge distribution models on the evolution of electric fields and beam parameters within 100 ps after the laser interaction with the target.

The impact of space charge increases with beam charge. We simulated different numbers of protons per pulse between $10^8$ and $10^{12}$ to study the effects, using the waterbag distribution model.

The transverse and longitudinal electric fields over 100 ps for different values of total charge are shown in Figs. 4 (a) and (b) respectively. They all decay quickly with the expansion of the beam.

The divergence angles and envelopes are shown in Figs. 4 (c) and (d) respectively. It is clear that drift of the beam will be affected dramatically if the total effective charge exceeds $10^{10}$ protons per pulse; meanwhile, energy spread and lengths also increase remarkably, as shown in Fig. 4 (e) and (f) respectively. This increase of divergence angle and energy spread will damage the beam quality. The influence of the space charge force diminishes greatly after 20 ps, corresponding to 1.2 mm from the target, even in the case of $10^{12}$ protons per pulse. The reason for this is that higher charge densities lead to more intense expansion, which further leads to a faster decrease in charge density.

As the proton numbers are all smaller than electron numbers in Table 1, the influence of the space charge force can be neglected during this initial drift. In a transport beamline, the electrons will be moved out of the beam under the effect of electromagnetic field elements. The elements (such as a quadrupole lens for focusing and collecting) in a beamline are usually placed not less than 0.1 m from the target; space charge force can be neglected.

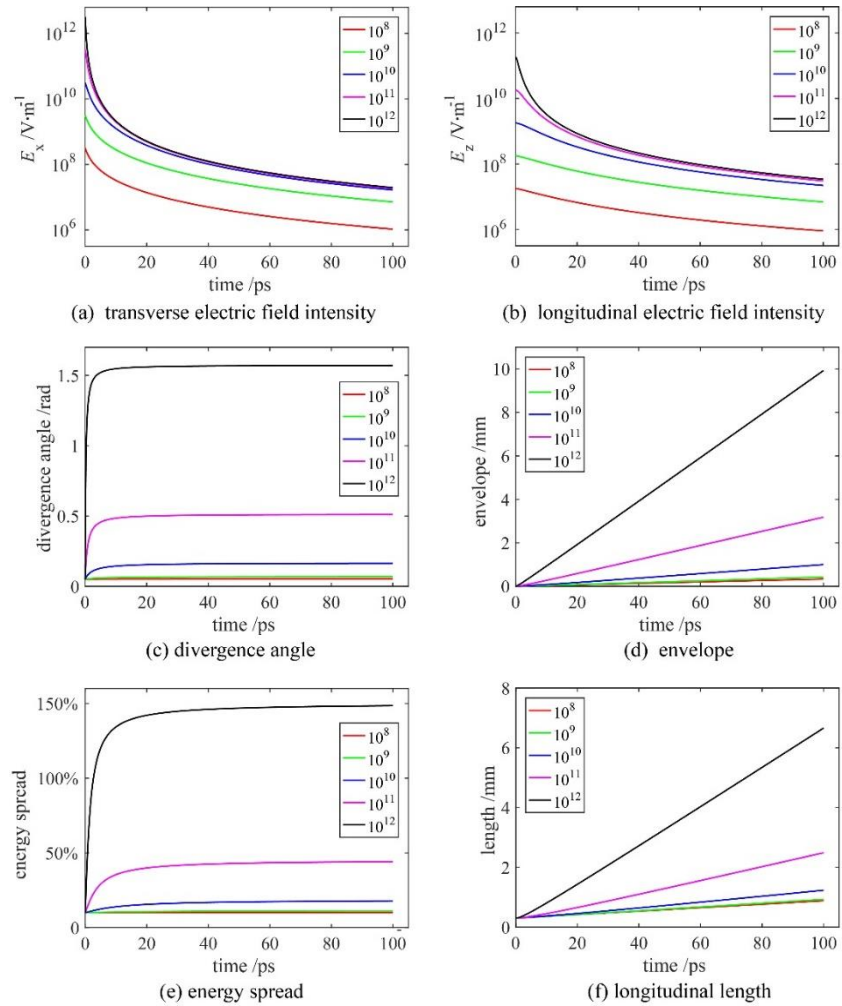

Fig. 4   Influence of space charge for different values of total charge on the evolution of electric fields and beam parameters within 100 ps after the laser interaction with the target.

If the beam can be transmitted using point-to-point optics, a proton beam with high charge and without electrons will be impacted by the space charge force at the exit. For protons and electrons with the same velocities, the bending radius of a proton is much larger than that of an electron. Hence an electron beam can be introduced for neutralizing space charge, and manipulated by magnetic field elements without influence on the proton beam at the exit.

For electron beams with an energy of 100 MeV, energy spread of ±1%, divergence angle of ±5 mrad, pulse duration of 100 fs and 20 μm radius of beam spot, if the number of electrons per pulse exceeds $10^5$, space charge force will have remarkable influence.

## 3 Conclusion

In summary, in order to study the influence of the co-moving electrons, we performed a series of two-dimensional PIC simulations with different targets (aluminum, plastic, DLC). We found that, with an energy spread of ±10% and initial divergence angles of ±50 mrad, the electron numbers are dominating and the space charge force of the protons can be neutralized by electrons after acceleration. Ellipsoid models with different charge distributions (linear-descent, waterbag, uniform) were used to compute the space charge force. In the linear-descent distribution, the influence of the transverse electric field was greater, and in the uniform distribution, the longitudinal electric field had a greater effect. In the waterbag distribution model, if the number of protons per pulse exceeds $10^{10}$, energy spread and divergence angle will increase dramatically and damage the beam quality. The influence of the space charge force diminish greatly after 20 ps, corresponding to 1.2 mm from the target, even in the case of $10^{12}$ protons per pulse. If the high-current proton beam can be transmitted using point-to-point optics, electrons will be moved out of the beam and would need to be introduced for neutralizing at the exit.

**Acknowledgements**

The authors are very grateful to Matthew J Easton for spending much time and energy making careful revisions to our paper, and the help from Dongyu Li, Kun Zhu, Xueqing Yan and Chen Lin at Peking University. The EPOCH code is used under UK EPSRC contract (EP/G055165/1 and EP/G056803/1).

# 激光加速质子束空间电荷力研究

朱军高[1]，赵　媛[1]，赖美福[1]，古永力[1]，徐世祥[2]，周沧涛[1]，卢海洋[1]

（1. 深圳技术大学 工程物理学院，先进材料测试技术研究中心，深圳市超强激光与先进材料技术重点实验室，广东 深圳 518118；2. 深圳大学 物理与光电工程学院，深圳市微纳光子信息技术重点实验室，教育部/广东省共建光电子器件和系统重点实验室，广东 深圳 518060）

**摘　要**：激光加速器可以输出具有独特品质的质子束，例如 μm 尺寸、ps 脉冲长度和高峰值电流。强流粒子束的空间电荷力效应较强，对面向应用的束流传输提出了挑战。通过二维 PIC 模拟研究了激光加速后与质子速度接近的电子的影响。采用椭球模型估算空间电荷力的影响，比较不同电荷分布的差异。结果表明每束团质子数超过 $10^{10}$ 后空间电荷力显著影响质子束传输，甚至严重破坏束流品质。空间电荷力的影响在 20 ps 后显著减弱，离开靶约 1.2 mm。

**关键词**：激光加速；质子束；空间电荷力；高亮度